\documentclass[a4paper,12pt,final]{article}
\usepackage{t1enc}
\usepackage[english]{babel}
\usepackage[utf8]{inputenc}
\usepackage{graphicx}
\usepackage{psfrag}
\usepackage{bbm}
\usepackage{amsmath}
\usepackage{amssymb}
\usepackage{showkeys}
\usepackage{hyperref}
\usepackage{ifthen}
\usepackage{epstopdf}

\title{``Half a proton'' in the Bogomol'nyi-Prasad-Sommerfield Skyrme model}
\author{Árpád Lukács\\
{\small MTA Wigner RCP RMI, Budapest, P.O.B.\ 49, H1525 Hungary}
}

\def\d{\mathrm{d}}

\def\lag{{\mathcal{L}}}

\def\kihagy#1{}

\def\Tr{\ensuremath{\mathop{\rm Tr}}}

\newcommand{\arxiv}[2][]{
  \ifthenelse{\equal{#1}{}}{
    \href{http://arxiv.org/abs/#2}{\texttt{arXiv:#2}}
  }{
    \href{http://arxiv.org/abs/#2}{\texttt{arXiv:#2 [#1]}}
  } 
}

\begin{document}
\maketitle

\begin{abstract}
The BPS Skyrme model is a model containing an $SU(2)$-valued scalar field, in which a Bogomol'nyi-type inequality can be satisfied by soliton solutions (skyrmions).
In this model, the energy density of static configurations is the sum of the square of the topological charge density plus a potential. The topological charge density
is nothing else but the pull-back of the Haar measure of the group $SU(2)$ on the physical space by the field configuration. As a consequence, this energy expression
has a high degree of symmetry: it is invariant to volume preserving diffeomorphisms both on physical space and on the target space. We demonstrate here, that in the BPS Skyrme model
such solutions exists, that a fraction of their charge and energy densities are localised, and the remaining part can be any far away, not interacting with the localised part.
\end{abstract}

\section{The model considered}\label{sec:model}
The Skyrme model \cite{Skyrme1, Skyrme2, Skyrme3} is a solitonic model of nuclear matter. It contains an $SU(2)$ valued scalar field, $U$, identified with the pion field,
and baryons and nuclei emerge in it as topological solitons, called skyrmions. For a review, see Ref.\ \cite{SMrev}. In Ref.\ \cite{Adam1, Adam2, Adam3}, a variant of the model has been put forward and termed the BPS Skyrme model,
in which solutions satisfying a Bogomol'nyi type bound are admitted \cite{Bogomolny, deVegaSchap1}. The model has been used to describe baryons \cite{Adam1, Adam2, Adam3, Adam4}, nuclei \cite{Adam5}, and, with coupling to gravity, neutron stars \cite{Adam6, Adam7}.

In the present paper, the BPS Skyrme model is considered, and it is demonstrated that due to its high degree of symmetry, the model also possesses some remarkable features: with somewhat less restrictive boundary conditions, solutions with finite energy but non-integer charge are obtained. Note, however, that this property is expected to be lost in perturbed models, therefore it does not affect the phenomenological applicability of the model. We would also like to mention, that in non-BPS and baby variants of the Skyrme model, states that 
can be 
considered bound states (or molecules) of skyrmions with fractional
charge have been obtained in Ref.\ \cite{GudnasonNitta, JSS, KobayashiNitta}. For fractional-charge configurations in $U(1)$ gauge theory with regular field-strength on a two-sphere,
see Refs.\ \cite{Kiskis1, Kiskis2}, and for fractionally charged self-dual solutions in an $SU(2)$ gauge theory, Ref.\ \cite{FHP}.

The BPS Skyrme model is defined with the action
\begin{equation}
  \label{eq:lag}
  S = \int \d^4 x \lag = \int \d^4 x \left[
    -\lambda^2 \pi^2 B^\mu B_\mu -\mu^2 V  \right]\,,
\end{equation}
where $B^\mu$ is the topological current density, defined as
\begin{equation}
  \label{eq:topcurr}
  B^\mu = \frac{1}{24\pi^2}\epsilon^{\mu\nu\rho\sigma}\Tr (L_\nu L_\rho L_\sigma)\,,\quad L_\mu = U^\dagger \partial_\mu U\,.
\end{equation}
Here $U$ is the pion field, taking its value in the group $SU(2)$ (the target space), and $\epsilon^{\mu\nu\rho\sigma}$ is the four-dimensional Levi-Civitá tensor.

The potential $V$ is assumed to be a function of $\Tr U$ only, e.g., $V = (1- (\Tr U)/2)^a$ with $a$ a positive real parameter (e.g., $a=1,2,3,\dots$).
The constants $\lambda$ and $\mu$ can be scaled away by suitable rescalings of the coordinates and fields  \cite{Adam1, Adam2, Adam3, Adam4, IL}.

Let us first calculate the time-like component of the topological current, the topological charge density,
\begin{equation}
 \label{eq:tchg}
 B^0 = -\frac{1}{24\pi^2}\epsilon_{ijk}\Tr (L_i L_j L_k)=\frac{1}{8\pi^2}\left( \Tr(L_3 L_2 L_1) - \Tr(L_1 L_2 L_3)\right)\,,
\end{equation}
where $\epsilon_{ijk}$ is the three-dimensional Levi-Civitá tensor.
The charge density in Eq.\ (\ref{eq:tchg}) is the pull-back of the invariant volume form of the group $SU(2)$, therefore, it is invariant to all volume preserving diffeomorphisms on both the physical and on the target space \cite{Adam1, Adam2, Adam3}.

For static configurations, all but the time-like component of $B^\mu$ vanish, and the energy density calculated from the action (\ref{eq:lag}) is
\begin{equation}
 \label{eq:erg}
 E=\int \d^3 x \mathcal{E} = \int \d^3 x \left[ \lambda^2\pi^2 (B^0)^2 + \mu^2 V \right]\,,
\end{equation}
which, as a consequence of the invariance of the charge, is also invariant to volume preserving diffeomorphisms both on physical and target space.


\section{Skyrmions}\label{sec:skyrmions}
Let us now consider spherically symmetric BPS skyrmion, i.e., soliton solutions of the BPS Skyrme model. A simple Ansatz for spherical symmetry is obtained as follows \cite{Adam1, Adam2, Adam3}.
Let
\begin{equation}
  \label{eq:param}
  U = \exp( i f {\bf \sigma}\hat{\bf n} )\,,
\end{equation}
where ${\bf \sigma}=(\sigma^1,\sigma^2,\sigma^3)$ denotes the Pauli matrices, and ${\bf n}$ is a unit vector field,
\begin{equation}
 \label{eq:nvec}
 \hat{\bf n} = (\sin\vartheta\cos n\varphi, \sin\vartheta\sin n\varphi, \cos\vartheta)\,,
\end{equation}
where $0\le r <\infty,0\le \vartheta \le \pi, -\pi < \varphi \le \pi$ denote the usual spherical coordinates in space, and $n$ is an integer. For such configurations, it is easily calculated, that
the only non-zero component of the topological current density is
\begin{equation}
 \label{eq:sphcurr}
 B^0 = -\frac{n}{2\pi^2 r^2} \sin^2 f(r) f'(r)\,,
\end{equation}
and thus if $f(0)=\pi$, $f(r\to\infty)=0$, the winding number $\int \d^3 x B^0$ agrees with $n$. For the energy density, we get
\begin{equation}
  \label{eq:BPS_erg}
  E = 4\pi \int\d r r^2 \left[ \left\{ \frac{n\lambda\sin^2 f}{2\pi r^2}f' \pm  \mu\sqrt{V}\right\}^2 \mp \frac{n\lambda\mu\sqrt{V}\sin^2 f}{\pi r^2}f'\right]\,.
\end{equation}
The last term is topological, and thus, in each topological sector a minimum is obtained if the
\begin{equation}
  \label{eq:BPS_eq}
  f' = \mp \frac{2\pi\mu \sqrt{V}r^2}{n\lambda\sin^2 f}
\end{equation}
Bogomol'nyi-Prasad-Sommerfield (BPS) equations hold, i.e., the term in the curly bracket in Eq.\ (\ref{eq:BPS_erg}) vanishes.

Solutions of the BPS equation can be found numerically \cite{IL}. In the $a=1$ case, a closed form solution is also known \cite{Adam1}. Note also,
that solutions of the radial equation (\ref{eq:BPS_eq}) with different winding number $n$ are related to each other via rescalings \cite{Adam1}.
(Note, that there is a sharp contrast with the original Skyrme model of Refs.\ \cite{Skyrme1, Skyrme2, Skyrme3}, where skyrmions have definite shape, and higher winding number skyrmions are not spherically symmetric.)

\section{Solutions with fractional charge}\label{sec:arbchg}
Let us now use the invariance of the model against volume preserving diffeomorphisms to construct solutions, where a fraction of the charge is in a large volume,
and the rest is far away, not interacting.

\begin{figure}
 \noindent\hfil\includegraphics{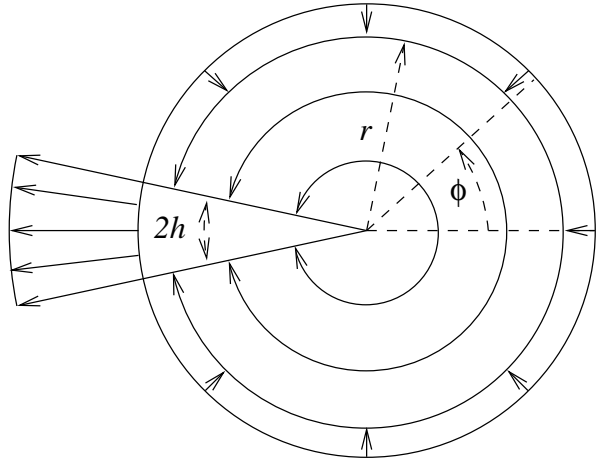}
 \caption{The effect of the transformation on the skyrmion (in the $\vartheta=\pi/2$ plane): a part containing charge $n$ is being deformed into a fractionally charged skyrmion, and the slice containing the remaining part, $\nu$, is squeezed out to larger distances.}
 \label{fig:radtrf}
\end{figure}

Intuitively, what we are going to do, is to take a spherically symmetric skyrmion with the charge rounded up to the next integer.
We then keep a part containing the desired amount of charge, and remove a slice containing the rest. This is done by making the angle of the slice thinner, squeezing the slice farther out, and deforming the remaining part into a smaller sphere (see Fig.\ \ref{fig:radtrf}).
As a function of the parameter determining how thin the charge containing the excess charge is, one can imagine the action of the transformations as the motion of a fluid.

Let $U_n(r,\vartheta, \varphi)$ denote the spherically symmetric solution obtained in Section \ref{sec:skyrmions}, and consider, as our new solution, $U_N(\tilde{r}, \vartheta, \tilde{\varphi})$, with
\begin{equation}\label{eq:phitilde}
 N\tilde{\varphi} = \left\{ \begin{aligned}
  n \varphi &\text{, if } |\varphi| < \pi - h\\
  n \varphi + \frac{\nu\pi}{h^2}(\varphi - \pi + h)^2 &\text{, if } \varphi > \pi -h\\
  -\tilde{\varphi}(-\varphi) &\text{, if } \varphi < 0
  \end{aligned}\right.\quad\quad
  \tilde{r} = ({\tilde{\varphi}}'(\varphi))^{-1/3} r\,,
\end{equation}
where $h >0$ is a small number, $N$ is the integer obtained by rounding $n$ \emph{away from zero} and $\nu = N-n$ (if necessary, a smooth function
approximating $\tilde{\varphi}$ may be used). Due to the invariance of the energy density, the result is still a static solution of the field equations. 

\begin{figure}
 \noindent\hfil\includegraphics[scale=.5,angle=-90]{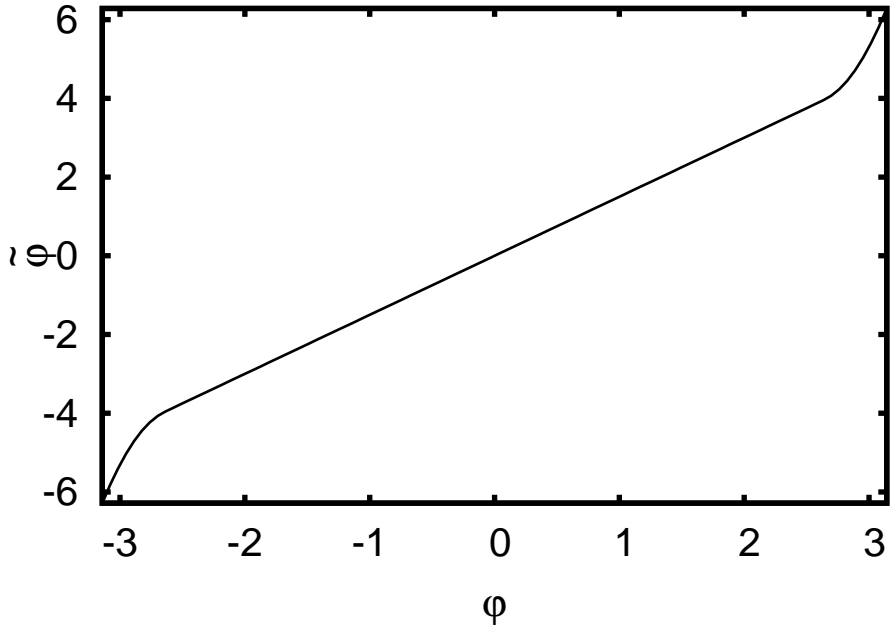}
 \caption{The function $\tilde{\varphi}$, determining the angle dependence of the fractionally charged skyrmion, shown here for $n=1.5$ and $h=0.5$.}
 \label{fig:angle}
\end{figure}

The function $\tilde\varphi$ is plotted in Fig.\ \ref{fig:angle}, and the effect of the transformation on the skyrmion is sketched
in Fig.\ \ref{fig:radtrf}. The points inside the skyrmion, at angles $|\varphi| < \pi - h$ are moved towards the origin, and to a larger angle. At the same time, the points with angular coordinate $|\varphi| > \pi - h$ are moved closer to the $\varphi=\pi$ line, and away from the origin.

The charge energy densities in the model [see Eqs.\ (\ref{eq:tchg}) and (\ref{eq:erg})] are invariant against volume preserving diffeomorphisms, therefore, the charge and energy densities are also only moved along with the points, with the transformation (\ref{eq:phitilde}), and their integrals are unchanged.

The charge in the region $|\varphi| < \pi-h$ is $n$ for the new solution. The remaining part, $\nu=N-n$, in
the region $|\varphi| > \pi-h$, is being pushed away from the origin, farther for smaller $h$. In this region, the
charge and energy densities are bounded by their respective values in the spherically symmetric solution at the origin.

As $h$ approaches zero, in a finite volume, the $|\varphi| > \pi-h$ region becomes very thin, and, as the energy and charge densities here are bounded, its total energy and
charge become arbitrarily small. In this sense, it becomes unobservable. As a result, we get a solution, where only a fraction of the charge is localised in
an arbitrarily large volume, not interacting with the remaining part far away.

For $h\to 0$ the point-wise limit of our configuration is also given by the Ansatz (\ref{eq:param}), however, with $n$ real (not necessarily integer).
In this sense, we have constructed a skyrmion with fractional charge. This field configuration jumps at $\varphi = \pm \pi$.

\section{Conclusions}
In the present paper, we have demonstrated, that the BPS Skyrme model admits solutions with a non-integer charge localised in a volume that can be taken of
any size, not interacting with the remaining part. The localised part of the solution is spherically symmetric apart from a thin band connecting it to the far away part,
and approaches a fractional solution in the usual Ansatz point-wise. The possibility of such solutions is a consequence of the high degree of symmetry of the model.
Using the symmetries of the model, the band connecting the fractional skyrmion with its remaining part can be made arbitrarily thin, containing a negligible amount of charge and
energy.


In the BPS Skyrme model, unlike for, e.g., a free scalar field, there is no energy term quadratic in the first derivatives of the field. As a result of this unusual property of the model, large changes in the field variable over small distances can give small contribution to the energy.
In other words, this is a consequence of the topological nature of the charge and that the kinetic energy is the charge squared: the energy is insensitive to distances, as far as volumes are preserved.

One of the motivations of the BPS Skyrme model is, that it describes a fluid \cite{Adam1, Adam2, Adam3}. Adding terms which give a surface tension to the fluid would make the energy
of the bands connecting the fractional skyrmion parts proportional to their area, in a sense, leading to a confining potential between them.

\subsection*{Acknowledgements}
This work has been supported by grant OTKA K101709. The author would like to thank P.~Forgács and M.~Horváth for reading the manuscript.

\end{document}